\documentclass[12pt,draftcls,onecolumn]{IEEEtran}

\IEEEoverridecommandlockouts                              

\usepackage{cite}
\usepackage{enumerate}
\usepackage[english]{babel}
\usepackage[utf8]{inputenc}
\usepackage{algorithm}
\usepackage[noend]{algpseudocode}
\usepackage{amsmath}
\usepackage{amssymb}
\usepackage{amsthm}
\usepackage{graphicx}
\usepackage{hyperref}
\usepackage{graphics}
\usepackage{algpseudocode}
\usepackage{subcaption}
\usepackage{tikz}
\usepackage{pgfplots}
\usetikzlibrary{shapes,decorations,shadows}  
\usetikzlibrary{decorations.pathmorphing}  
\usetikzlibrary{decorations.shapes}  
\usetikzlibrary{fadings}
\usetikzlibrary{patterns}
\usetikzlibrary{arrows,shapes}
\usetikzlibrary{calc}
\usetikzlibrary{positioning} 
\tikzset{
    >=stealth',
    punkt/.style={
           rectangle,
           rounded corners,
           draw=black, thick,
           text width=10.5em,
           minimum height=2em,
           text centered,
           },
    pil/.style={
           ->,
           thick,
           shorten <=2pt,
           shorten >=2pt,}
}

\usepackage{amssymb}
\usepackage{amsthm}

\theoremstyle{plain}

\theoremstyle{plain}

\theoremstyle{plain}

\newtheorem{remark}{Remark}

\title{Actor-Critic Algorithms for Constrained Multi-agent Reinforcement Learning \footnote{A version of this paper appeared as an extended abstract in Proceedings of the 18th International Conference on Autonomous Agents and Multiagent Systems
(AAMAS 2019) \cite{diddigi2019actor}. Equal contribution by the first three authors}}

\author{
Raghuram Bharadwaj Diddigi$^{1}$
Danda Saikoti Reddy$^{2}$
Prabuchandran K.J.$^{1}$ 
Shalabh Bhatnagar$^1$

$^1$ Department of Computer Science and Automation, IISc Bangalore, India\\
$^2$ IBM Research, Banaglore, India\\

\{raghub, prabuchandra,shalabh\}@iisc.ac.in,
saikotireddy@in.ibm.com
}

\begin{document}
\maketitle
\thispagestyle{empty}
\pagestyle{empty}
\begin{abstract}
 
 In cooperative stochastic games multiple agents work towards learning joint optimal actions in an unknown environment to achieve a common goal. In many real-world applications, however, constraints are often imposed on the actions that can be jointly taken by the agents. In such scenarios the agents aim to learn joint actions to achieve a common goal (minimizing a specified cost function) while meeting the given constraints (specified via certain penalty functions). In this paper, we consider the relaxation of the constrained optimization problem by constructing the Lagrangian of the cost and penalty functions. We propose a nested actor-critic solution approach to solve this relaxed problem. In this approach, an actor-critic scheme is employed to improve the policy for a given Lagrange parameter update on a faster timescale as in the classical actor-critic architecture. A meta actor-critic scheme using this faster timescale policy updates is then employed to improve the Lagrange parameters on the slower timescale. Utilizing the proposed nested actor-critic schemes,  we develop three Nested Actor-Critic (N-AC) algorithms. Through experiments on constrained cooperative tasks, we show the effectiveness of the proposed algorithms.
 
\end{abstract}

\section{Introduction}
In the reinforcement learning (RL) paradigm, an agent interacts with its environment by selecting actions in a trial and error manner. The agent incurs cost for the chosen actions and the goal of the agent is to learn to choose actions to  minimize a long-run cost objective. The evolution of the state of the environment and the cost feedback signal received by the agent is modeled using the standard Markov Decision Process (MDP) \cite{sutton1998introduction} framework. Utilizing one of the RL methods like Q-learning \cite{sutton1998introduction}, the agent learns to choose optimal state dependent actions (policy) by suitably balancing exploration of unexplored actions and exploiting the actions that incur low long-run costs. However, in many problems of practical interest the number of environment states and the set of actions that the agent has to explore for learning the optimal actions are typically high resulting in the phenomenon `curse of dimensionality'. In such high-dimensional scenarios, RL methods in conjunction with deep neural networks as function approximators known as ``critic-only'' or ``actor-critic'' methods have resulted in successful practical applications \cite{mnih2015human,levine2016end}.




Many real world problems nonetheless cannot be considered in the context of single agent RL and has led to the study of multi-agent RL framework  \cite{marlsurvey}. It is important to observe that developing learning methods in the multi-agent setting poses a serious challenge compared to the single agent setting due to the exponential growth in state and action spaces as the number of agents increase. 

Multi-agent reinforcement learning problems have been posed and studied in the mathematical framework of ``stochastic games''  \cite{lauer2000algorithm}.  The stochastic game setting could be categorized as (a) fully cooperative \cite{littman2001value,lauer2000algorithm,foerster2017counterfactual}, (b) fully competitive \cite{littman1994markov} and  (c) mixed  settings \cite{lowe2017multi,foerster2018learning}. In a fully cooperative game, agents coordinate with other agents either through explicit communication \cite{das2017learning,foerster2016learning,mordatch2017emergence} or through their actions to achieve a common goal. In this paper, we consider the fully cooperative setting which has gained popularity in recent times \cite{foerster2017stabilising,gupta2017cooperative}.

In many real-life multi-agent applications one often encounters constraints specified on the sequence of actions taken by the agents. Under this setting, the combined goal of the agents is to obtain the optimal joint action sequence or policy that minimizes a long-run objective function while meeting the constraints that are typically specified as long-run penalty/budget functional constraints. It is important to observe that both the objective as well as the penalty functions depend on the joint policy of the agents. These problems are studied as ``Constrained Markov Decision Process'' (C-MDP) \cite{altman1999constrained} for the single agent RL setting and as a ``Constrained Stochastic Game (C-SG)'' for the multi-agent RL settings. 


In this work, our goal is to develop multi-agent RL algorithms for the setting of constrained cooperative stochastic games. To this end, we utilize the Lagrange formulation and propose novel actor-critic algorithms. Our algorithms, in addition to the classical actor-critic setup, utilize an additional meta actor-critic architecture to enforce constraints on the agents. The meta actor performs gradient ascent on the Lagrange parameters by obtaining the gradient information from the meta critic. We propose three RL algorithms namely JAL N-AC, Independent N-AC and Centralized N-AC by extending three popular algorithms of the unconstrained cooperative SGs to the constrained fully cooperative SGs. We now summarize our contributions:
\begin{itemize}
    \item We propose, for the first time, multi-agent actor-critic algorithms for the constrained fully cooperative stochastic game setting. 
    \item Our algorithms do not require model information to be known and utilize non-linear function approximation in the actor as well as critic for modeling the policy as well as value function.
    \item We utilize a meta actor-critic architecture in addition to the classical actor-critic setup to satisfy the specified constraints. 
    \item The meta critic utilizes a non-linear function approximator for obtaining the value function of the penalty costs. 
    \item Under this setup, we develop three RL algorithms for the constrained multi-agent setting. 
    \item We provide empirical evaluation of the performance of our algorithms on certain constrained multi-agent tasks. 
\end{itemize}


\subsection{Related Work}
The long-run average cost as the objective function with long-run average cost constraints for the single agent MDP setting has been considered in \cite{borkar2005actor} and a two-time scale actor-critic scheme utilizing full state representation without function approximation has been developed under this C-MDP setting. In \cite{lakshmanan2012novel} and  \cite{bhatnagar2010actor,bhatnagar2012online} a constrained Q-learning algorithm as well as actor-critic algorithms, respectively, utilizing linear function approximators  have been proposed. Recently deep neural network based value function approximators for C-MDPs under the discounted cost objective setting have been presented in \cite{achiam2017constrained} and a constrained policy optimization (CPO) algorithm has been developed for the continuous C-MDPs for near constraint satisfaction. 

\section{Model}\label{model}
We first consider the problem of obtaining joint optimal action sequence in the cooperative multi-agent setting. This problem can be formulated in the framework of stochastic games. A stochastic game is an extension of the single agent Markov Decision Process to multiple agents. A stochastic game is described by the tuple $(n,S,A_{1},...A_{n},T,C,\gamma)$ where $n$ denotes the number of agents participating in the game, $S$ denotes the state space of the game, $A_i, ~ i \in{1, \ldots, n}$ denotes the action space of the agents, $C: S \times A_{1} \times...\times A_{n}\times S \xrightarrow{} \mathbb{R}$ denotes the common cost function for the cost incurred by the agents when the joint action profile is $(a_1,a_2,\ldots,a_n), ~ a_i \in A_i, ~ i \in \{1,2 \ldots, n\}$, $T: S \times A_{1} \times...\times A_{n} \times S \xrightarrow{}[0,1] $ denotes the probability transition mechanism where $T(i,a_1,\ldots,a_n,j)$  specifies the probability of transitioning to state $j$ from the current state $i$ under the  joint action profile  $(a_{1},...a_{n})$ of the agents and $\gamma \in (0,1]$ is the discount factor.

Let $X_{t} \in S$ denote the state of the game at time $t$. Assume that the initial state $X_{0}$ is sampled from an initial distribution $D$. Let $\pi_{i} : S \times A \xrightarrow{} [0,1]$ be the stochastic policy followed by the agent $i$. Here $\pi_i(a|s)$ for agent $i$ specifies the probability of choosing action $a \in A_i$ in state $s$. Given a joint policy of the agents $\pi = (\pi_1,\ldots \pi_n)$, we define the total discounted cost incurred for the joint policy as 
\begin{align}\label{unconstrained}
  J(\pi) =  E \Big[\sum_{t=0}^{\tau-1} \gamma^{t}C(X_{t},\pi(X_{t}),X_{t+1})\Big],
\end{align}
where $E[\cdot]$ denotes the expectation taken over the sequence of states under the joint policy $\pi$, $\tau$ denotes the number of time steps until the terminal state is reached in the game (random but finite integer).

The objective of the agents in the cooperative stochastic game is to learn a joint optimal policy $\pi^{*} = (\pi^{*}_1,\ldots,\pi^{*}_n)$ that minimizes \eqref{unconstrained}, i.e., 
\begin{align}\label{optimal}
    \pi^{*} = \arg \min_{\pi} J(\pi).
\end{align}

In the constrained cooperative stochastic game setting, we consider $K$ common total discounted penalty constraints with single stage cost functions $P_{j}:S\times A_{1} \times...\times A_{n}\times S \xrightarrow{} \mathbb{R}, ~j \in \{1,\ldots, K\}$. These constraints are specified as \begin{align}\label{constrained}
  E \Big[\sum_{t=0}^{\tau -1} \gamma^{t} P_{j}(X_{t},\pi(X_{t}),X_{t+1})\Big] \leq \alpha_{j}, ~j \in \{1,\ldots, K\},
\end{align}
where $\alpha_j \geq 0, ~ j \in 1,\ldots K$ are certain prescribed thresholds. Under this constrained stochastic game setting, the objective of the agents is to learn a joint policy $\pi$ that minimizes \eqref{unconstrained} under  constraints \eqref{constrained}.

In order to solve for $\pi^*$ in \eqref{optimal} subject to the constraints \eqref{constrained}, we consider the Lagrangian formulation of the multi-agent constrained setting \cite{borkar2005actor,bhatnagar2010actor}.
Let $\lambda_j, ~j \in \{1,\ldots,K\}$ denote the Lagrange multipliers for each of these constraints. Let $\lambda = (\lambda_{1}, \ldots, \lambda_{K})$ denote the vector of Lagrange multipliers. We define the Lagrangian cost function as follows:
\begin{align}\label{lagrangian}
L(\pi,\lambda) &=  E \Big[\sum_{t=0}^{\tau -1} \gamma^{t} \big( C(X_{t},\pi(X_{t}),X_{t+1}) ~+ \\
& \nonumber \sum_{i=1}^{K} \lambda_{j} P_{j}(X_{t},\pi(X_{t}),X_{t+1}) \big) \Big]-\sum_{j=1}^{K} \lambda_{j} \alpha_{j}.
\end{align}
Let $g: \mathbb{R}^K \xrightarrow{} \mathbb{R}$ denote the dual objective of the constrained problem that is defined as
\begin{align} \label{lag-cost}
    g(\lambda) = \inf_\pi L(\pi,\lambda).
\end{align}
For a given vector of Lagrange multipliers, $g(\lambda)$ can be computed by optimally solving  the unconstrained MDP with the modified single stage cost function $\tilde{C}$ given by
\begin{align}\label{lag-single-state-cost}
\tilde{C}(X_{t},\pi(X_{t}),X_{t+1}) &=    C(X_{t},\pi(X_{t}),X_{t+1}) + \nonumber \\
& \sum_{j=1}^{K} \lambda_{j} P_{j}(X_{t},\pi(X_{t}),X_{t+1}).
\end{align}

Let $q: \mathbb{R}^K \xrightarrow{} \Pi$ denote the optimal policy obtained by solving the unconstrained problem with the modified cost function \eqref{lag-single-state-cost}, i.e.,
\begin{align}\label{find-policy}
    q(\lambda) = \arg \min_{\pi \in \Pi}L(\pi,\lambda),
\end{align}
where $\Pi$ denotes the space of all joint randomized policies. 

After constructing the dual of the constrained problem, the goal then is to maximize $g(\lambda)$ with respect to Lagrange multipliers $\lambda$. Let $\lambda^*$ denote the optimal Lagrange multipliers obtained by maximizing g($\lambda$), i.e.,
\begin{align}
    \lambda^* = \arg \max_{\lambda \geq 0} g(\lambda).
\end{align}
The optimal Lagrange multipliers $\lambda^*$ can be obtained by performing gradient ascent on the function $g(\lambda)$, i.e.,
\begin{align}\label{find-lm}
    \lambda_{t+1} = (\lambda_{t} + b(t) \nabla g(\lambda))^+ ,
\end{align}
where $b(t),~ t \geq 0$ is a suitably chosen step-size schedule and $(\cdot)^+$ denotes the function $\max(\cdot,0)$. 
The gradient of $g(\lambda)$ with respect to the Lagrange multipliers $\lambda_j, ~j \in \{1,\ldots,K\}$ can be obtained using the envelope theorem of mathematical economics as follows (see \cite{borkar2005actor}):

\begin{align}\label{gradObj}
    \frac{\partial g(\lambda)}{\partial \lambda_j} &= \frac{\partial L(\pi,\lambda)}{\partial \lambda_j}\bigg|_{\pi=q(\lambda)}, ~~j \in \{1,2,\ldots,K\} \nonumber\\
    &= E \Big[\sum_{t=0}^{\tau -1} \gamma^t P_{j}(X_{t},q(\lambda)(X_{t}),X_{t+1})\Big] - \alpha_{j}.
\end{align}
Equation \eqref{gradObj} indicates that the partial derivative with respect to the Lagrange multipliers $\lambda_j, ~j \in \{1,2,\ldots,K\}$ can be computed by performing policy evaluations of the policy $q(\lambda_t)$ corresponding to single-stage cost functions $P_{j}(X_{t},q(\lambda_t)(X_{t}),X_{t+1})$. The policy $q(\lambda^*)$ corresponding to the $\lambda^*$ at the end of the gradient iterations provides a near-optimal policy satisfying \eqref{constrained}, i.e.,
\begin{align}
    \pi^* = q(\lambda^*).
\end{align}

%

\begin{figure}[!h]
\begin{center}
\scalebox{0.8}{
\begin{tikzpicture}[node distance=1cm, auto]
 \node[punkt] (critic) {Policy Critic \\ (Estimates the value function for the modified cost MDP) };
 \node[punkt,below=1cm of critic] (actor){Policy Actor \\ (Improves the policy)};
 \node[punkt,below=1.5cm of actor] (pcritic) {Penalty Critic \\ (Estimates the value function of penalty cost for the policy $q(\lambda)$};
 \node[punkt,above=1.5cm of critic] (pactor) {Penalty Actor \\ (Improves Lagrange multipliers)};
 

 \draw[-, very thick,draw=black!80] (critic.east) to  ($(critic.east)+(2,0)$);
 \draw[->, very thick,draw=black!80] ($(actor.east)+(2,0)$) to (actor.east);
 \draw[-, very thick,draw=black!80] ($(critic.east)+(2,0)$) to ($(actor.east)+(2,0)$);
 \draw[-,very thick,draw=black!80] (actor.west) to ($(actor.west)+(-2,0)$);
 \draw[-,very thick,draw=black!80] ($(actor.west)+(-2,0)$) to ($(critic.west)+(-2,0)$);
 \draw[->,very thick,draw=black!80] ($(critic.west)+(-2,0)$) to ($(critic.west)$);
 
 \draw[-, very thick,draw=black!80] (pcritic.east) to ($(pcritic.east)+(2.5,0)$);
 \draw[->, very thick,draw=black!80] ($(pactor.east)+(2.5,0)$) to node [] {$\nabla g(\lambda)$}  (pactor.east);
 \draw[-, very thick,draw=black!80] ($(pcritic.east)+(2.5,0)$) to ($(pactor.east)+(2.5,0)$);
 \draw[-,very thick,draw=black!80] (pactor.west) to ($(pactor.west)+(-2.5,0)$);
 \draw[-,very thick,draw=black!80] ($(pactor.west)+(-2.5,0)$) to ($(pcritic.west)+(-2.5,0)$);
 \draw[->,very thick,draw=black!80] ($(pcritic.west)+(-2.5,0)$) to ($(pcritic.west)$);

 \draw[red,very thick, dotted] ($(critic.north west)+(-1.0,0.6)$)  rectangle ($(actor.south east)+(1.0,-0.6)$);
 \draw[red,very thick, dotted] ($(pactor.north west)+(-3.0,0.6)$)  rectangle ($(pcritic.south east)+(3.0,-0.6)$);
 \draw[->,black,thick] ($(pactor.south)$)  to node [right] {$\lambda$} ($(critic.north)$);
 \draw[->,black,thick] ($(actor.south)$)  to node [right] {$q(\lambda)$} ($(pcritic.north)$);
 \end{tikzpicture}
 }
\end{center}
\caption{Nested Actor-Critic (N-AC) architecture}
\label{doubleac}
\end{figure}
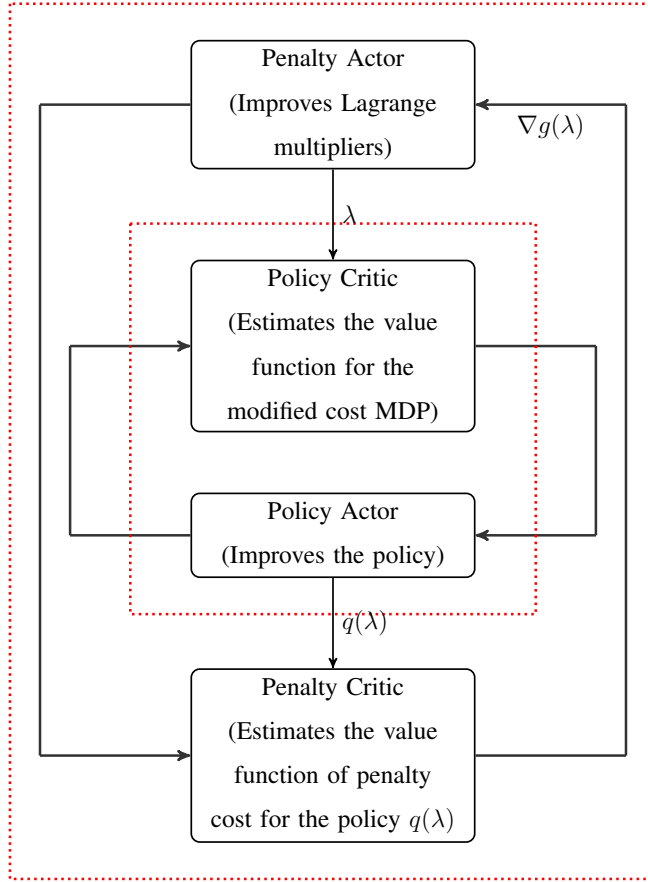

To accomplish the task of computing gradients by policy evaluation \eqref{gradObj} and improving Lagrange multipliers \eqref{find-lm}, we propose nested actor-critic architectures as illustrated in Figure \ref{doubleac}. In this setup, the inner (policy) actor-critic computes the optimal policy that minimizes \eqref{lagrangian} (utilizing \eqref{lag-single-state-cost}) for a fixed set of Lagrange multipliers $\lambda$. The policy obtained from the inner actor-critic is given as input to the outer (penalty) critic. The penalty critic then computes the gradient with respect to Lagrangian by evaluating the policy $q(\lambda)$ for all the penalty functions as in \eqref{gradObj}. The gradient information is then provided to the outer (penalty) actor to improve the Lagrange multipliers by performing gradient ascent as in \eqref{find-lm}. Finally, the outer actor provides the improved Lagrange multipliers to the policy actor-critic for obtaining the optimal policy at the improved Lagrange multipliers.


\section{Proposed Algorithms}\label{algo}
In this section, we propose three multi-agent deep reinforcement learning algorithms in the constrained setting. 
First, we propose the Joint Action Learners (JAL) scheme for the constrained case and refer to this algorithm as ``JAL N-AC''. This algorithm employs  centralized critics and a centralized actors. As the number of agents that participate  in the game increase, the learning becomes slow due to action explosion. Next, to mitigate the action explosion of JAL N-AC, we propose independent learners scheme for the constrained setting and refer to this algorithm as ``Independent N-AC ''.  In Independent N-AC, both critics and actors are decentralized. Note that even though this algorithm handles action explosion, decentralizing critics induces non-stationarity to the learning process for each of the agents. Finally, we also propose an actor-critic algorithm that employs ``centralized learning and decentralized execution" where there is a single policy and penalty critic and multiple policy actors, and refer to this algorithm as ``Centralized N-AC''. 

\subsection{JAL N-AC}
JAL N-AC employs centralized critics and centralized actors for all the agents. The centralized policy actor computes the joint policy of all the agents in the game. Therefore, the action space of the centralized policy actor is the cartesian product of action spaces of all the agents in the game.

We will now describe the JAL N-AC algorithm. Let us denote the current sample of the game at time $t$ by the tuple $(X_{t},a_{t}, X_{t+1},C_{t},\{P_{1_{t}},\ldots,P_{K_{t}}\})$, where $X_{t}$ is the current state, $a_{t}$ is the joint action taken by the central policy actor, $X_{t+1}$ is the next state, $C_{t}$ is the single-stage cost obtained from the environment, and $\{P_{1_{t}},\ldots,P_{K_{t}}\}$ are $K$ single-stage penalty costs. Let $\theta_c$, $\theta_\pi$ and $\theta_{p_{j}},~ j \in \{1,\ldots,K\}$ correspond to the parameters of the policy critic, policy actor and penalty critic respectively. 
Policy critic parameters $\theta_{c}$ are updated by minimizing the loss function \cite{mnih2015human},
\begin{align}
    L(\theta_{c}) = (r_{t} + \gamma V_{\theta_{c}}(X_{t+1}) - V_{\theta_{c}}(X_{t}))^{2},
\end{align}
where $r_t$ is the modified single-stage cost given by 
$r_{t} = C(X_{t},a_{t},X_{t+1}) + \sum_{j=1}^{K}\lambda_{j} P_{j_{t}}(X_{t},a_{t},X_{t+1}$~
and $V_{\theta_{c}}(\cdot)$ denotes the value function approximated by the policy critic.

Having found the parameters $\theta_c$ of the policy critic, we utilize it to compute policy gradients for improving the policy parameters $\theta_{\pi}$ of the actor. There are many ways to estimate the gradient for improving the actor parameters \cite{sutton1998introduction}. We utilize the popular temporal difference learning (TD(0)) update with baseline. We update the policy parameters $\theta_{\pi}$ as follows:
\begin{align}\label{dac-policy}
   \theta_{\pi} := \theta_{\pi} - a(t)  (& r_{t} + \gamma V_{\theta_{c}}(X_{t+1}) - V_{\theta_{c}}(X_{t})) \nonumber \\ & \nabla_{\theta_{\pi}} \log \pi(a_{t}|X_{t})),
\end{align}
where $V_{\theta_{c}}(X_{t})$ is the baseline. Note that in \eqref{dac-policy} the baseline is subtracted from the value function estimate to reduce the variance of the gradient estimate.  

The penalty critic estimates the penalty value function parameters $\theta_{p_{j}} , j \in 1,\ldots,K$. These parameters are computed by minimizing the loss function $L(\theta_{p_j})$ defined as 
\begin{align}
    L(\theta_{p_{j}}) = (P_{j}(X_{t},a_{t},X_{t+1}) +& \gamma V_{\theta_{p_{j}}}(X_{t+1}) - V_{\theta_{p_{j}}}(X_{t}))^{2}, \nonumber
\end{align}
where $V_{\theta_{p_{j}}}(X_{t})$ (resp. $V_{\theta_{p_{j}}}(X_{t+1})$) is the value function associated with the penalty constraint $j$ for state $X_{t}$ (resp. $X_{t+1}$).

Finally, the Lagrange parameters are improved by the penalty actor by performing stochastic gradient ascent as follows (see  \cite{borkar2005actor,bhatnagar2010actor}):
\begin{align}\label{dac-lag}
    \lambda_{j_{t+1}} = max(0, \lambda_{j_{t}} + b(t)(V_{\theta_{p_j}}(X_{t}) - \alpha_{j})),
\end{align}
where $b(t)$ is the step-size parameter and $\lambda_{j_{t}}$ is the Lagrange parameter corresponding to penalty function $i$ at time $t$. The maximum operation is done to ensure that the Lagrange parameters stay always positive. The update in \eqref{dac-policy} is performed on a faster time scale while the update in \eqref{dac-lag} is performed on a slower timescale \cite{borkar2005actor}.


\subsection{Independent N-AC}
In this algorithm, each agent has its own nested actor-critic architecture, i.e., there are a total of $n$ nested actor-critic architectures. Each agent learns parameters separately for its nested actor-critic architecture and estimates its individual policy $\pi_{i}$, i.e., each agent maintains its own policy actor-critic and penalty actor-critic networks. At every step of training, each agent takes actions based on its current policy independent of other agents policies and receives common cost from the environment. Using this cost signal, all the agents independently improve their policy and penalty parameters. Each agent manages its nested actor-critic architecture in the same manner as described in the JAL N-AC algorithm. Independent N-AC suffers from the problem of non-stationarity  as past learning of an agent may become obsolete as other agents simultaneously explore actions during their training phase. Therefore the individual policies obtained by the agents may not be optimal. Nonetheless this is a simple algorithm that avoids action explosion and has been seen to perform well in some scenarios \cite{tampuu2017multiagent}. 

\subsection{Centralized N-AC}
This algorithm imbibes advantages of the two algorithms described above. During training, learning is centralized here in the sense that value function is estimated based on the joint actions of all agents while policies for all agents are decentralized. There is one centralized critic (policy critic) for estimating the value function of the single state cost (i.e., the Lagrangian), another centralized critic (penalty critic) for estimating the value function of the penalty cost (for improving the Lagrange parameters) and $n$ actors estimating the policies of each of the agents. Finally, there is a penalty actor improving the Lagrange multipliers. After learning is complete, agents execute learnt policies independently. This idea is well studied in the unconstrained multi-agent case in \cite{foerster2017counterfactual,lowe2017multi}. 
The algorithmic description of Centralized N-AC for solving the fully cooperative multi-agent constrained RL problem is provided in Algorithm \ref{C-AC}.

\begin{remark}
In our algorithms, policy actor-critic determines the optimal policy by minimizing the Lagrangian for a given $\lambda$. On the other hand, the penalty actor-critic updates the Lagrange multipliers by evaluating the policy on the penalty cost functions. As these two computations have to be carried ad infinitum, the idea of two time-scale stochastic approximations \cite{borkar1997stochastic} has been utilized to interleave these two operations for ensuring the desired convergence behaviour. 
\end{remark}
\begin{algorithm}
\caption{Centralized N-AC}
\label{C-AC}
\begin{algorithmic}[1]
    \State State sample at time $t$:  $X_{t} = (X_{t}^{1},...X_{t}^{i})$ where $X_{t}^{i}$ is the state of the agent $i$.
    \For{agents $i = 1,2,\ldots,n$}
    \State $a_{i}$ = Sample an action from $\pi_{i}(\cdot \mid X_{t}^{i})$
    \EndFor
    \State Obtain cost, penalties and next state from the environment.$$ (C_{t},P_{j_{t}},X_{t+1}) \xleftarrow{} get\_reward(X_{t},a_{1},...a_{n}) $$
    \State Let $r_{t} = C_{t} + \sum_{j=1}^{K} \lambda_{j_t} P_{j_{t}}.$
	\State Train policy critic parameters $\theta_{c}$ to minimize the loss function $(r_{t} + \gamma V_{\theta_{c}}(X_{t+1}) - V_{\theta_{c}}(X_{t}))^{2}$
	\For{$j = 1,2,\ldots,K$}
	\State Train penalty critic parameters $\theta_{p_{j}}$ to minimize the loss function  $(P_{j_{t}} + \gamma V_{\theta_{p_{j}}}(X_{t+1}) - V_{\theta_{p_{j}}}(X_{t}))^{2}$
	\EndFor
	\For{agents $i = 1,\ldots,n$}
	\State Improve policy actor $i$'s policy parameter $\theta_{\pi_{i}}$ by performing gradient descent along the estimated gradient  $(r_{t} + \gamma V_{\theta_{c}}(X_{t+1}) - V_{\theta_{c}}(X_{t})) \nabla_{\theta_{\pi_{i}}} \log \pi_{i}(a_{i} \mid X_{t}^{i})$
	\EndFor
	\State Finally update the Lagrange parameters in the penalty actor as
	$\lambda_{t+1} = \max(0, \lambda_t + b(t)(V_{\theta_{p}}(X_{t}) - \alpha))$
	where $\alpha=(\alpha_1,\ldots,\alpha_K)$ is the vector of prescribed thresholds.
	\end{algorithmic}
\end{algorithm}

\section{Experiments and Results}\label{exp}
In this section, we evaluate and analyze our algorithms on three constrained muti-agent tasks. We begin with two simple games namely constrained grid world and constrained coin game that have discrete state spaces. We then discuss our results on a complex environment - constrained cooperative navigation that has continuous state space. 

\subsection{Constrained Grid World}
In the constrained grid world, the objective of each agent is to learn the shortest path from a given source to the target with at most $\alpha$ overlap in the path with other agents. For our setting, we consider a grid of size $4\times 4$ with two agents. The state of each agent $s_i, i \in 1,2$ is a vector of size 16 with value 1 at the current position of the agent $i$ and value 0 at all other positions. The permissible actions for agents in the grid include moving up, down, left and right wherever applicable. The game ends when both the agents reach the target state $11$ or the number of steps in the game exceeds 10. Note that when an agent reaches the target state, it remains in the target state till the end of the episode. In the constrained setting that we consider, a single-stage penalty of +1 is imposed on the agents if they enter the same block in the grid and we prescribe a penalty threshold of $\alpha$. For example, if we let $\alpha$ to be 0, then we are imposing the constraint that the agents have to reach the target state from every source state in minimum number of steps without any overlap in their paths.

\begin{table}[h]
\centering
\begin{tabular}{|l|l|l|l|}
\hline
12 & 13 & 14 & 15          \\ \hline
8  & 9  & 10 & \textbf{11} \\ \hline
4  & 5  & 6  & 7           \\ \hline
0  & 1  & 2  & 3           \\ \hline
\end{tabular}
\caption{Grid World}
\label{GW-fig}
\end{table}


In this experiment, we train all three algorithms for $10,000$ episodes starting from random start positions and three different $\alpha$ values $0.1, 0.3$ and $0.5$ respectively. We perform $10$ independent runs of the experiment and report the median of the expected penalty obtained across $10$ runs. Note that the expected penalty is computed by averaging total penalty obtained by following the converged policy over $10,000$ test episodes. We observe that in all  the three algorithms, agents reach the target state from any given start state in at most 10 steps. The performance of our algorithms in meeting the constraints is given in Table \ref{GW-conv} and we find that all algorithms nearly meet the penalty constraints. 

We now briefly discuss how the agents learn the shortest path to reach the target state while meeting the penalty constraints. For example, two agents starting from state $0$ learn to take the following paths to reach $11$, the target state. $$ Agent 1: 0-4-8-9-10-11$$ and $$ Agent 2: 0-1-2-6-7-11.$$ On the other hand, for the initial state $3$, both the agents learn to follow the same path: $$ 3-7-11.$$ In the first case (start state 0), agents took disjoint paths to reach the target state in the least number of steps. In the second case (start state 3), however, if one of the agents takes a detour from the shortest route to reach the target state, then it considerably increases the objective of the game. Hence, they learn to take the same shortest route violating the constraint minimally when required. Note that average overlap is 1 when we start from state 3 however as the initial state is chosen with probability 1/16, the average overlap is 0.06. 

In Table \ref{GW-steps}, we present the median of the expected cost for three distinct values of $\alpha = 0.1,0.3,0.5$ obtained by our algorithms. Note that the expected cost in this experiment is the average number of steps taken by the algorithm to reach the target state from random start positions. The expected cost monotonically decreases with increase in $\alpha$. This is the desired behaviour as the constraint becomes less tighter when $\alpha$ increases and is seen to be the case in all our algorithms. 

\begin{table}[]
\begin{center}
\begin{tabular}{|c|c|c|c|}
\hline
\textbf{Algorithm}                                          & \textbf{\begin{tabular}[c]{@{}l@{}}Expected \\ Penalty \\ ($\alpha = 0.1$)\end{tabular}} & \textbf{\begin{tabular}[c]{@{}l@{}}Expected \\ Penalty \\  ($\alpha = 0.3$)\end{tabular}} & \textbf{\begin{tabular}[c]{@{}l@{}}Expected \\ Penalty \\ ($\alpha = 0.5$)\end{tabular}} \\ \hline
\begin{tabular}[c]{@{}l@{}}JAL\\ 
N-AC\end{tabular} & 0.092                                                                                 & 0.250                                                                                  & 0.444                                                                              \\ \hline

\begin{tabular}[c]{@{}l@{}}Independent\\  N-AC\end{tabular} & 0.127                                                                                 & 0.221                                                                                  & 0.346                                                                                 \\ \hline
\begin{tabular}[c]{@{}l@{}}Centralised \\ N-AC\end{tabular} & 0.064                                                                                 & 0.217                                                                                  & 0.405                                                                                 \\ \hline
\end{tabular}
\caption{Expected penalty obtained by the converged policy for different values of penalty threshold in constrained grid world}
\label{GW-conv}
\end{center}
\end{table}


\begin{table}[]
\centering
\begin{tabular}{|c|c|c|c|}
\hline
\textbf{Algorithm}                                          & \textbf{\begin{tabular}[c]{@{}c@{}}Expected \\ Total Cost\\ $\alpha = 0.1$\end{tabular}} & \textbf{\begin{tabular}[c]{@{}c@{}}Expected\\ Total Cost\\ $\alpha = 0.3$\end{tabular}} & \textbf{\begin{tabular}[c]{@{}c@{}}Expected\\ Total Cost\\ $\alpha = 0.5$\end{tabular}} \\ \hline
\begin{tabular}[c]{@{}c@{}}JAL \\ N-AC\end{tabular}         & 2.4149                                                                                   & 2.1837                                                                                  & 2.1831                                                                                  \\ \hline
\begin{tabular}[c]{@{}c@{}}Independent\\  N-AC\end{tabular} & 1.8144                                                                                        & 1.6276                                                                                       & 1.5594                                                                                       \\ \hline
\begin{tabular}[c]{@{}c@{}}Centralised \\ N-AC\end{tabular} & 2.1457                                                                                   & 1.7104                                                                                  & 1.5607                                                                                  \\ \hline
\end{tabular}
\caption{Expected total cost of the converged policy in the constrained grid world for different thresholds}
\label{GW-steps}
\end{table}

\subsection{Constrained Coin Game}
In the coin game considered in \cite{foerster2018learning}, the objective of the agents is to collect the coin that appear at random positions in the given grid. In the constrained version, multiple agents exist and each agent can only collect specific type of coin. The objective is to maximize the total coins collected by the agents. We consider two agents `blue' and `red' in a $3 \times 3$ grid. The coin can be in one of the two colors - blue or red. We impose a penalty on the agents if the color of the agent doesn't match with the color of the coin collected. For example, if the agent 'blue' collects a 'red' coin, a penalty of +1 is incurred by both the agents. The state of the game is a $4 \times 3 \times 3$ matrix that encodes the positions of the agents in the grid and also positions of the coins (blue and red) \cite{foerster2018learning}. Note that unlike in the grid world game, both agents have access to full state information. The actions of agents similar to the grid world setting include moving up, down, left and right wherever applicable.

\begin{table}[h]
\begin{center}

\begin{tabular}{|c|c|}
\hline
\textbf{Algorithm} & \multicolumn{1}{l|}{\textbf{\begin{tabular}[c]{@{}l@{}}Expected Penalty \\ ~~~\quad $\alpha = 0.2 $\end{tabular}}} \\ \hline
JAL N-AC           & 0.110                                                                                                         \\ \hline
Independent N-AC   & 0.211                                                                                                          \\ \hline
Centralized N-AC   & 0.208                                                                                                           \\ \hline
\end{tabular}
\caption{Performance of converged policy in meeting the constraints in the constrained coin game}
\label{CG-conv}
\end{center}
\end{table}

We evaluate the performance of the converged policy across $10$ runs and report the median of the expected penalty in Table \ref{CG-conv}. We observe that all three algorithms nearly meet the penalty threshold value $\alpha = 0.2$. 

\subsection{Constrained Cooperative Navigation}
This is the constrained version of the cooperative navigation game proposed in \cite{lowe2017multi,mordatch2017emergence}. The objective of the agents in this game is to move towards the landmarks that are located on a continuous space. Note that this game despite having similarity to constrained grid world has continuous state space unlike the finite state space in the grid world. As we have uncountable number of states in this game, non-linear function approximators for estimating the cost and penalty value functions play a crucial role in obtaining good policy. The positions of the agents and landmarks change dynamically over different episodes. The agents have to learn a policy that minimizes the number of steps to reach the landmark with constraint on the number of collisions between the agents. For each landmark, the Euclidean distance to the closest agent is calculated and sum of the distances is provided as the common cost to the agents. Each agent incurs a penalty of +1 for colliding with each other. 


\begin{figure}[h!]
\begin{center}
\includegraphics[scale = 0.5]{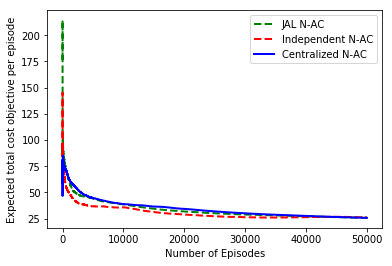}
\caption{Performance of algorithms in reducing the objective function as the learning progresses}
\label{cc_lag}
\end{center}
\end{figure}

\begin{figure}[h!]
\begin{center}
\includegraphics[scale = 0.5]{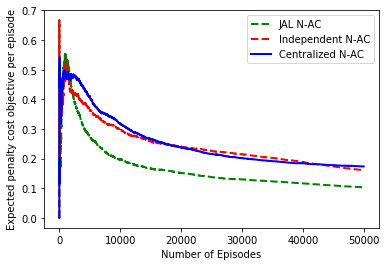}
\caption{Performance of algorithms in meeting the constraint as the learning progresses}
\label{cc_pen}
\end{center}
\end{figure}

\begin{table}[]
\begin{center}
\begin{tabular}{|c|c|c|}
\hline
\textbf{Algorithm} & \textbf{\begin{tabular}[c]{@{}c@{}}Expected \\ Total Cost \\ $\alpha = 0.1$\end{tabular}} & \textbf{\begin{tabular}[c]{@{}c@{}}Expected \\ Penalty\\ $\alpha = 0.1$\end{tabular}} \\ \hline
JAL N-AC & 18.5826           & 0.0442                                                                                                                    \\ \hline
Independent N-AC  & 22.9310 & 0.0338                                                                                                                    \\ \hline
Centralized N-AC   & 17.9201 & 0.1324                                                                                                                    \\ \hline
\end{tabular}
\end{center}
\caption{Performance of converged policy in constrained cooperative navigation}
\label{ss-conv}
\end{table}

In this experiment, we have two agents in the game. The agent is said to reach the landmark if the total Euclidean distance cost is less than $2$ units. The game also ends if agents do not reach the landmark in maximum of $30$ time steps. We set the penalty constraint to $\alpha = 0.1$. We train our algorithm on a single run for $5 \times 10^{4}$ iterations. From Figures \ref{cc_lag} and \ref{cc_pen}, 
we see that the expected total cost and expected penalty decreases as learning progresses. 
Finally, in Table \ref{ss-conv}, we observe that all our algorithms yield converged policies that nearly satisfy the penalty constraints.

\section{Conclusions and Future Work}\label{conc}
In this paper, we have considered the problem of finding near-optimal policies satisfying specified constraints for the multi-agent fully cooperative stochastic game setting. Our algorithms utilize nested actor-critic architectures to enforce agents to meet the penalty constraints. Utilizing this architecture, we presented three multi-agent RL methods namely JAL N-AC, Independent N-AC and Centralized N-AC each of which utilize non-linear function approximators for value function estimations. 
Finally, we empirically showed the performance of our algorithms on three multi-agent tasks.

An interesting future direction would be to extend the proposed actor-critic algorithms to other constrained stochastic games involving say fully competitive and mixed (cooperative and competitive) settings. Another line of research would be to develop algorithms for agents with continuous action spaces. Further, we would like to deploy our algorithms on real world applications such as cooperative surveillance through multiple drones and smart power grid settings with constraints. 

\clearpage
\bibliographystyle{IEEEtran}
\bibliography{sample-bibliography}

\end{document}


\maketitle

\section{Convergence Analysis}

In this section, we present the convergence analysis of centralized N-AC algorithm for the constrained cooperative stochastic games. We show that the algorithm converges to locally optimal policy which satisfies the constraints. The following are the assumptions needed for the convergence of the algorithm:
\begin{asu} \label{asu1}
    The underlying  Markov chain for any agent $i$ determining the dynamics $\{X^{\theta_i}_n\}$, under stationary random policy $\pi^{\theta_i}$, is irreducible and aperiodic.
\end{asu}
\begin{asu} \label{asu2}
$\pi(\textbf{a}|s) = \pi_{1}(a_{1}|s).\pi_{2}(a_{2}|s)\ldots \pi_{n}(a_{n}|s).$
\end{asu}
\begin{asu} \label{asu3}
    The algorithm uses the unbiased estimate of the value function $V^{\pi}$ for parameter update of the policy. 
\end{asu}
\begin{asu} \label{asu4}
     The policy $\pi^{\theta_i}(s,a)$ is continuously differentiable in $\theta_i$.
\end{asu}
\begin{asu} \label{asu5}
The step sizes $a^{n}, b^{n}, c^{n}$ for all $n \geq 0$ satisfy the standard stochastic approximation algorithm conditions:
 \begin{equation}\label{sastepcond}
\begin{aligned}
& \Sigma_{n} a_{n} = \Sigma_{n} b_{n} = \Sigma_{n} c_{n} = \infty\\
& \Sigma_{n} (a_{n})^2 , \Sigma_{n} (b_{n})^2 , \Sigma_{n} (c_{n})^2 < \infty \\
&  a_n = o (b_n), b = o (c_n).
\end{aligned}
\end{equation}
\end{asu}
\begin{remark}[Unbiased estimate of $V^{\pi}$]
The unbiased estimate of $V^{\pi}$ can be obtained  for full state representation and for compatible linear function approximation in single agent RL. However, the ``centralized critic" architecture allows the estimate of $V^{\pi}$ to be unbiased in multi-agent RL as well.
\end{remark}

\begin{theorem}
Under the Assumptions $\ref{asu1}-\ref{asu5}$, the sequence of policy parameter updates $\theta^{n}_{i}$ for an agent $i$ 
converges to $\theta^{*}_{i}$ almost surely. Here, $\theta^{*}_{i}$ is a locally optimal policy parameter and $\pi^{\theta^{*}_i}$ satisfies the respective  constraints.
\end{theorem}
\begin{proof}
We use the ordinary differential equations (ODE) approach for stochastic approximation (SA) algorithms to prove the asymptotic convergence. The high level overview of the proof technique is as follows:
\begin{step}[Two-time scale SA convergence \cite{borkar1997stochastic}]
We show that the updates of parameters $(\theta_{i},\lambda_{i})$ converges at different rates to the stationary point $(\theta^{*}_i,\lambda^*_i)$ almost surely.
\end{step}
\begin{step}[Lyapunov stability] We show that the Lagrangian $L(\theta_i,\lambda_i)$ is indeed the Lyapunov function in order to prove that the stationary point $(\theta^{*}_i,\lambda^*_i)$ is locally asymptotically stable.
\end{step}
\begin{step} [Saddle point theorem]
We use the saddle point theorem \cite{bertsekas1999nonlinear} to conclude that the $\theta^{*}_i$ of the stationary point $(\theta^{*}_i,\lambda^*_i)$ is the local optima for constraint cooperative stochastic game.
\end{step}
\begin{remark}[Step 1]
We use the Two time scale SA scheme, i.e, $\theta_i$ converges faster compared to  $\lambda_i$ due to its faster time scale. 
\end{remark}
The parameters which are in slower time scale are invariant to the parameter updates on faster time scale. While analyzing $\theta_i$ update the parameter $\lambda_i$ is fixed. One can construct the ODE corresponding to the respective parameter updates to show that they converges to the stationary point $(\theta^{*}_i,\lambda^*_i)$.
\begin{remark}[Step2]
The Lyapunov functions to show that the iterates are asymptotically stable are as follows: 
\begin{align*}
    \mathbb{L}_{\lambda_i}(\theta_i) = L(\theta_i, \lambda_i) - L(\theta^{*}_i, \lambda_i) \,\,\,\text{for}\,\,\, \theta_i
\end{align*} and 
\begin{align*}
\mathbb{L}(\lambda_i) = - L(\theta^{*}_i, \lambda_i) + L(\theta^{*}_i, \lambda^{*}_i) \,\,\,\text{for}\,\,\, \lambda_i
\end{align*}
Here, $\theta^{*}_i$ is local minimum where as $\lambda^{*}_i$ is local maximum point.
\end{remark}
We show that the sequence $\{\theta^{n}_i\}$ converges to the local minimum of $L(\theta_i,\lambda_i)$ for a fixed $\lambda_i$. Also, we show that $\lambda_i$ sequence converges to local maximum point.
\begin{remark}[Step3]
We show that $(\theta^{*}_i,\lambda^*_i)$ is a local saddle point of the Lagrangian $L(\theta_i,\lambda_i)$. By saddle point theorem $\theta^{*}_i$ is a local optimal solution for constrained cooperative stochastic game.
\end{remark}
\end{proof}

\bibliographystyle{named}
\bibliography{sample-bibliography}